\documentclass{article}
\usepackage{spconf,amsmath,graphicx, cite}

\title{Dual Knowledge Distillation for Efficient Sound Event Detection}
\name{Yang Xiao and Rohan Kumar Das}
\address{Fortemedia Singapore, Singapore\\ \small \{xiaoyang, rohankd\}@fortemedia.com}

\begin{document}
%
\maketitle
\begin{abstract}
    Sound event detection (SED) is essential for recognizing specific sounds and their temporal locations within acoustic signals. This becomes challenging particularly for on-device applications, where computational resources are limited. To address this issue, we introduce a novel framework referred to as dual knowledge distillation for developing efficient SED systems in this work. Our proposed dual knowledge distillation commences with temporal-averaging knowledge distillation (TAKD), utilizing a mean student model derived from the temporal averaging of the student model's parameters. This allows the student model to indirectly learn from a pre-trained teacher model, ensuring a stable knowledge distillation. Subsequently, we introduce embedding-enhanced feature distillation (EEFD), which involves incorporating an embedding distillation layer within the student model to bolster contextual learning. On DCASE 2023 Task 4A public evaluation dataset, our proposed SED system with dual knowledge distillation having merely one-third of the baseline model's parameters, demonstrates superior performance in terms of PSDS1 and PSDS2. This highlights the importance of proposed dual knowledge distillation for compact SED systems, which can be ideal for edge devices.

\end{abstract}
\section{Introduction}
\label{sec:intro}

Sound event detection (SED)~\cite{sed} research has received attention in the field of audio signal processing, where the objective is to identify and temporally locate specific sound events within continuous audio streams. This technology has diverse applications, ranging from urban sound monitoring to automated surveillance, and from healthcare monitoring to environmental sound analysis~\cite{home, smarthome2}. A notable benchmark in this field is the detection and classification of acoustic scenes and events (DCASE) challenge series, which has one task to focus on the detection of sound events from domestic environments, emphasizing the need for models that can operate effectively in everyday settings. 

Deep learning advancements have greatly improved SED, particularly with convolution recurrent neural network (CRNN)~\cite{crnn,meanteacherbaseline3} models. A notable example is frequency dynamic (FDY)-CRNN~\cite{fdy,fmsg2}, which employs frequency-dependent kernels for better event detection but increases the model size by 150\%. In order to address SED model deployment on limited-capacity devices, lightweight attention mechanisms~\cite{secrnn} such as the squeeze-and-excitation (SE)~\cite{senet} module and time-frame frequency-wise SE (tfwSE) have been explored. However, these models with over 4 million parameters, are still not very suitable for edge devices. Additionally, using embeddings from large pre-trained models~\cite{pepe,pretrain} to enhance smaller ones faces practical challenges due to computational overhead and latency in on-device applications. Such challenges highlight the critical need for model compression in SED, ensuring efficient implementation in environments where resources are constrained. 

Knowledge distillation~\cite{kd} is a prevalent technique for model compression, which has been widely used in the audio processing applications such as in low-complexity acoustic scene classification for DCASE challenge~\cite{task1}. However, it has not been explored to that extent for SED applications. Some studies have introduced knowledge distillation~\cite{kdsed1,kdsed2,kdsed3} for SED by combining it with the mean teacher~\cite{meanteacherbaseline3, meanteacher,meanteacherbaseline1} model using semi-supervised learning. This approach enables effective training using unlabeled data. However, a critical observation in these implementations is that the mean teacher model and knowledge distillation often follow different optimization trajectories. This results into treating them as separate frameworks, which can limit the extent of knowledge transfer from the teacher to the student model.


In this study, we propose a novel dual knowledge distillation framework to address the challenges in developing on-device limited resource based SED systems. The proposed framework encompasses temporal-averaging knowledge distillation (TAKD), employing a mean student model derived from temporal averaging of parameters from the student model, enabling indirect learning from a pre-trained teacher model for stable distillation. Complementing this, we integrate embedding-enhanced feature distillation (EEFD), which adds a distillation layer to the student model to augment contextual learning. The proposed dual knowledge distillation framework comprising of TAKD and EEFD is evaluated on DCASE 2023 Task 4A for low-complexity efficient SED systems from the perspective of deploying them on edge devices.   

\section{Proposed Dual Knowledge Distillation }
\label{sec:methods}

\subsection{Backbone Model (SE-CRNN)}
\label{sec:se}

In our study, we adopt a modified CRNN, known as SE-CRNN~\cite{secrnn}, which incorporates an efficient attention mechanism for SED systems. The SE-CRNN model employs the squeeze-and-excitation (SE) method for channel-wise attention, refining the convolutional output. Additionally, it integrates a time-frame frequency-wise SE (tfwSE) module, applying frequency-wise attention to each time frame for a more powerful detection approach.

The SE-CRNN consists of seven convolution layers, each with SE and tfwSE modules except the final layer, leading up to two layers of bidirectional gated recurrent units (BiGRUs), and concluding with a fully connected layer. To address the high parameter count of the original SE-CRNN, we downsized the input to the convolution and BiGRU layers. This resulted in two variants: SE-CRNN-8 with [8, 16, 32, 64, 64, 64, 64] channels, and SE-CRNN-16 with [16, 32, 64, 128, 128, 128, 128] channels, with the GRU input matching the last convolution layer’s channel size. This streamlined architecture allows us to use compressed models as a reference to validate our proposed methods.


\subsection{Dual Knowledge Distillation}
\subsubsection{Temporal-Averaging Knowledge Distillation (TAKD)}
\label{takd}

In the recent SED research, the mean teacher semi-supervised method has been widely adopted in several studies~\cite{meanteacherbaseline3, meanteacherbaseline1, meanteacherbaseline2}. This method optimizes the use of unlabeled data by averaging model weights over training steps, which is generally more effective than using the final weights directly. Here, the teacher model, not partaking in backpropagation, employs the exponential moving average (EMA)~\cite{meanteacher} of the student model weights. The training involves calculating both classification and consistency losses, the latter encompassing class consistency at clip and frame levels, derived from comparing the logit of both student and teacher models across all audio clips.

In this study, we refer integration of knowledge distillation with the mean teacher method~\cite{kdsed1, kdsed2} as conventional dual-teacher distillation (CDTD), which involves a system with one student model and two teacher models: the `pre-trained teacher model' for knowledge distillation and the `mean teacher model' for the mean teacher method. 
CDTD introduces an additional term to the consistency loss for comparison between the logit of student and the pre-trained teacher model. However, this setup presents a challenge as the student model struggles with conflicting learning directions from the mean teacher, the pre-trained teacher, and the classification loss. This complexity can hinder the learning process, potentially leading to misalignment in objectives. During testing, while outputs from both the student and mean teacher models can be used for predictions, the predictions from mean teacher model are more reliable post-training. Therefore, relying on the student model for knowledge distillation may not be very effective in the mean teacher model.

In contrast, the proposed TAKD method offers a unique way to perform knowledge distillation. Similar to CDTD, the TAKD method uses a mean model derived from the temporal averaging of the weights from student model. The key distinction of TAKD lies in the integration of knowledge distillation, where it computes knowledge distillation loss between the mean model (referred to as the mean student) and the pre-trained teacher instead of the standard way of computing it between the student model and the pre-trained teacher as in CDTD. This enables the student model in TAKD to indirectly learn from the pre-trained teacher through the EMA-updated mean student. Thereby TAKD not only addresses conflicting teaching directions but also retains the enhanced performance characteristic of the mean student from CDTD. By redirecting the student model's learning focus, TAKD efficiently resolves the misalignment challenge inherent in CDTD, streamlining the learning process to develop effective SED models.



\begin{figure}[t]
\centering  
\includegraphics[width=0.65\columnwidth]{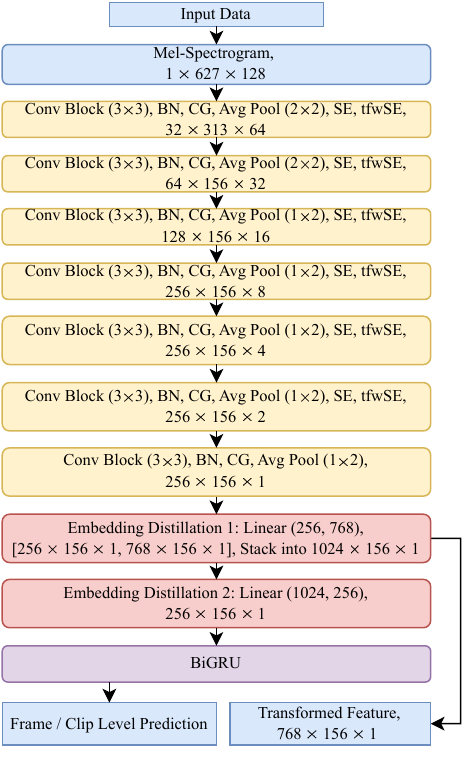}
\caption{SE-CRNN model with proposed EEFD. `BN' and `CG' denote batch normalization~\cite{bn} and the context gating~\cite{meanteacherbaseline1}, respectively. The shape of (x $\times$ y $\times$ z) indicates (channel $\times$ frame $\times$ frequency).
}
\label{fig:eefd}
\vspace{-4mm}
\end{figure}

\subsubsection{Embedding-Enhanced Feature Distillation (EEFD)}
\label{eefd}
The traditional approach to integrate pre-trained embeddings~\cite{pretrain} in SED involves extracting frame-level features from advanced models like BEATS~\cite{beats}, which are then merged with CRNN features through late fusion before the BiGRUs. This fusion process necessitates the use of aggregation techniques, notably adaptive average pooling, to ensure temporal dimension compatibility~\cite{pepe}. While the intent is to leverage the rich category-specific information from the embeddings, this integration significantly increases the computational load. The reliance on embeddings at inference time particularly intensifies the computational burden, making it less viable for on-device applications with stringent resource constraints. Therefore, this kind of approach may face practical challenges in real-time SED applications.

Our EEFD method innovatively integrates pre-trained embeddings into SED, circumventing the computational inefficiencies prevalent in native approaches. Instead of employing pre-trained embeddings at inference, which burdens computational resources, EEFD integrates these embeddings only during the training phase of the model. Fig.~\ref{fig:eefd} illustrates the SE-CRNN architecture with EEFD, outlining the layer configuration and showcasing our distillation strategy. This integration employs an `embedding distillation' mechanism, a linear layer that reshapes the post-convolution feature map to align with the first dimensions of the embeddings. The transformed features are then merged with the original convolution outputs and re-dimensioned before being processed by a BiGRU layer for advanced feature extraction. 

During training, the transformed features that aligned in the channel dimension undergo adaptive average pooling to attain temporal congruence, followed by a feature loss assessment against the pre-trained embeddings. This strategy preserves the contextual richness of the embeddings, enhancing the event classification accuracy of the model without incurring extra computational costs during inference. Consequently, EEFD upholds the performance of SED model, while ensuring its lightweight structure is maintained, perfectly fitting the resource constraints of on-device applications. 


\begin{figure}[t]
\centering  
\includegraphics[width=\columnwidth]{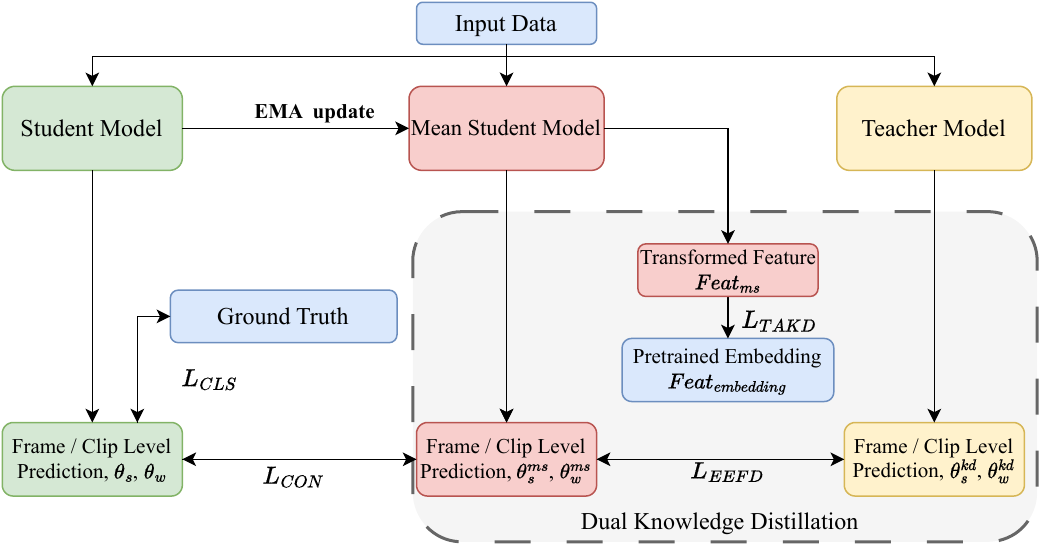}
\caption{SED system using dual knowledge distillation framework (highlighted by grey box). \(\theta_s\) and \(\theta_w\) represent the frame and clip level predictions of the student model, respectively.}
\label{fig:takd}
\vspace{-4mm}
\end{figure}


We now focus on the proposed {\it dual knowledge distillation} framework that combines our TAKD and EEFD methods, which are designed for developing efficient and lightweight SED systems.
Fig.~\ref{fig:takd} shows the proposed SED system using the dual knowledge distillation framework. At the start of each epoch, the student model processes input data to produce predictions \(\theta_s\) and \(\theta_w\) as shown in the Fig.~\ref{fig:takd}. These predictions are used to compare with ground truth labels to compute the classification loss \(L_{CLS}\) and compare with the mean student model's predictions \(\theta^{ms}_s\) and \(\theta^{ms}_w\) to determine the consistency loss \(L_{CON}\). Then, the mean student model, which does not undergo direct training, is updated using the EMA~\cite{meanteacher} of the student model's weights. In TAKD, the predictions from the mean student model are compared against those of the pre-trained teacher model to form the knowledge distillation loss \(L_{TAKD}\) as:
\vspace{-2mm}
\begin{equation}
\label{eq1}
    L_{TAKD} = \frac{1}{n} \sum_{i=1}^{n} (\theta^{ms}_w - \theta^{kd}_w)^2 + \frac{1}{n} \sum_{i=1}^{n} (\theta^{ms}_s - \theta^{kd}_s)^2
\vspace{-2mm}
\end{equation}
where \(\theta^{kd}_s\) and \(\theta^{kd}_w\) denote the pre-trained teacher model's frame and clip level predictions, respectively. 
Meanwhile, the mean student model outputs transformed features \(Feat_{ms}\) after embedding distillation layers. EEFD calculates a feature loss \(L_{EEFD}\) by measuring the difference between the transformed features and the pre-trained embeddings \(Feat_{embedding}\). The \(L_{EEFD}\) is defined as follows:
\vspace{-2mm}
\begin{equation}
\label{eq2}
    L_{EEFD} = \frac{1}{n} \sum_{i=1}^{n} (Feat_{embedding} - Feat_{ms})^2 
\vspace{-2mm}
\end{equation}
Therefore, the total loss \(L_{total}\) is derived by combining all the losses as follows:
\vspace{-1mm}
\begin{equation}
\label{eq3}
    L_{total} = L_{CLS} + \mu \times (L_{CON} + L_{TAKD} + L_{EEFD})
\vspace{-1mm}
\end{equation}
The symbol \(\mu\) represents the weight controlled by the scheduler. The student model's weights are updated through backpropagation, accounting for the \(L_{total}\). Simultaneously, the weights of the mean student model are softly updated to align with the student model's learning progression, ensuring a temporally smoothed knowledge transfer.

By integrating TAKD and EEFD, the dual knowledge distillation approach builds a more proficient student model and a stable mean student model with each epoch. It also maintains computational efficiency, making it well-suited for SED systems for on-device applications.

\section{Experiment Setting}
\label{sec:setting}
\subsection{Dataset and Evaluation Metric}
We considered the DCASE 2023 Task 4A dataset~\cite{desed}, featuring 10-second audio clips across various subsets. These include 1,578 weakly annotated real recordings, 14,412 unlabeled real recordings, 3,470 strongly labeled clips from AudioSet~\cite{audioset}, and 10,000 strongly annotated synthetic recordings~\cite{impact}. Performance evaluation was conducted on the public evaluation set. All clips were resampled to 16 kHz mono and segmented using a 2048-sample window and 256-sample hop length for spectrogram extraction and log-mel spectrogram generation.

Our systems were evaluated using the threshold-independent polyphonic sound event detection scores (PSDS)~\cite{tpsds,psds} in two scenarios following DCASE 2023 Task 4A protocol. Scenario-1 focuses on prompt reaction and temporal localization, while Scenario-2 emphasizes on reducing class confusion for SED.

\subsection{Implementation Details}

The experiments were conducted with a batch size of 48, using the Adam optimizer (learning rate: 0.001) and an exponential warmup scheduler for the first 50 of 200 epochs. The EMA factor of mean teacher was set to 0.999. To reduce complexity, we tested two SE-CRNN variants (SE-CRNN-8 and SE-CRNN-16) with reduced input sizes in convolution and BiGRU layers. The pre-trained BEATs model~\cite{beats} was used for frame-level embeddings of size 768.

\section{RESULTS AND ANALYSIS}
\label{sec:typestyle}

\begin{table}[t]
\begin{center}
\caption{Comparative results in PSDS1 and PSDS2 of the proposed TAKD and EEFD algorithms on the DCASE 2023 Task 4A public evaluation set. SE-CRNN-8 and SE-CRNN-16 are used as the student model for the studies.}
\vspace{2mm}
\label{table:skd}
\begin{tabular}{|c|c|c|}
\hline
\hfil \textbf{System} & \hfil \textbf{PSDS1} & \hfil \textbf{PSDS2}\\
\hline\hline
\hfil SE-CRNN (Pre-trained Teacher)  & \hfil 0.456 &\hfil 0.693\\
\hline 
\hfil SE-CRNN-8 & \hfil 0.347 &\hfil 0.572\\
\hline
\hfil SE-CRNN-16 & \hfil 0.401 & \hfil0.622\\
\hline \hline
\hfil SE-CRNN-8 + CDTD  & \hfil 0.344 & \hfil 0.582\\
\hline
\hfil SE-CRNN-16 + CDTD  & \hfil 0.402 & \hfil 0.614\\
\hline \hline
\hfil SE-CRNN-8 + TAKD  & \hfil 0.388 & \hfil0.607\\
\hline
\hfil SE-CRNN-16 + TAKD  & \hfil 0.438 & \hfil 0.662\\
\hline \hline
\hfil SE-CRNN-8 + EEFD  & \hfil 0.386 & \hfil0.603\\
\hline
\hfil SE-CRNN-16 + EEFD  & \hfil 0.442 & \hfil 0.676\\
\hline
\end{tabular}
\vspace{-4mm}
\end{center}
\end{table}

\subsection{Significance of the TAKD and EEFD}

We are first interested to evaluate the impact of both the knowledge distillation approaches used in our dual knowledge distillation framework. Table~\ref{table:skd} shows the comparative performance study for TAKD and EEFD methods with the pre-trained SE-CRNN teacher model, student models (SE-CRNN-8 and SE-CRNN-16) and CDTD based model, which is the conventional way of performing knowledge distillation. It is observed that TAKD based knowledge distillation performs much better than CDTD to improve the performance of the student models SE-CRNN-8 and SE-CRNN-16 to make it closer towards the performance of the teacher model. Similar to the TAKD method, the EEFD method also improves the performance of the two student models by using embedding as a feature representation.

\begin{table}[t]
\begin{center}
\caption{Performance comparison of our proposed SED system having dual knowledge distillation with
other reference models on the DCASE 2023 Task 4A public evaluation set. `Xiao-FMSG-task4a-1' is built upon our prior works~\cite{fmsg1,iscslp,ssp}.}
\vspace{-2mm}
\label{table:total}
\resizebox{\columnwidth}{!}
{%
\begin{tabular}{|c|c|c|c|}
\hline
\hfil \textbf{System} & \hfil \textbf{Param} & \hfil \textbf{PSDS1} & \hfil \textbf{PSDS2}\\
\hline\hline\
\hfil Barahona-AUDIAS-task4a-1 \cite{barahona}  & \hfil 1M & \hfil 0.427 & \hfil 0.625\\
\hline
\hfil Baseline CRNN \cite{tpsds}  & \hfil 1.2M & \hfil 0.366 &\hfil 0.575\\
\hline
\hfil Li-USTC-task4a-7 \cite{li}  & \hfil 2M & \hfil 0.451 & \hfil 0.673\\
\hline
\hfil Xiao-FMSG-task4a-1 \cite{fmsg1}  & \hfil 2.8M & \hfil 0.455 & \hfil0.695\\
\hline
\hfil SE-CRNN \cite{secrnn}  & \hfil 4.5M & \hfil 0.456 & \hfil0.693\\
\hline
\multicolumn{4}{|c|}{\textbf{Proposed Systems with Dual Knowledge Distillation}}\\
\hline
\hfil SE-CRNN-8 + EEFD + TAKD  & \hfil 400K & \hfil 0.407 & \hfil 0.628\\
+ AFL + DA & \hfil 400K & \hfil 0.428 & \hfil 0.636\\
\hline
\hfil SE-CRNN-16 + EEFD + TAKD  & \hfil 1.3M & \hfil 0.465 & \hfil 0.690\\
+ AFL + DA & \hfil 1.3M & \hfil 0.474 & \hfil 0.698\\
\hline
\end{tabular}%
}
\vspace{-4mm}
\end{center}
\end{table}

\subsection{Significance of Dual Knowledge Distillation}
We then evaluate the impact of our dual knowledge distillation framework on SE-CRNN-8 and SE-CRNN-16 with other reference models. Table~\ref{table:total} shows the comparative performance study with the pre-trained SE-CRNN teacher model, the baseline system in DCASE 2023 Task 4A challenge, and a few top submissions that are single models with similar model parameters. It is observed that our proposed dual knowledge distillation framework makes the SE-CRNN-8 to perform much better than the baseline CRNN even with one-third of the number of parameters. Similarly, the SE-CRNN-16 when integrated with our proposed framework shows a comparable performance to other top submissions, but with a comparatively less number of model parameters. We also employ the methods from our prior works to further enhance the SED system performance. We use the asymmetrical focal loss (AFL)~\cite{afl} as the classification loss and consider the data augmentation (DA) methods like mixup and time masking. Along with our proposed framework, it extends the performance of SE-CRNN-16 to the best in terms of PSDS1 and PSDS2 in comparison to all the systems discussed.

\section{CONCLUSION}
\label{sec:con}
In this work, we proposed a dual knowledge distillation framework, integrating TAKD and EEFD methods. This framework leverages the strengths of a mean student model for stability and an embedding layer for contextual learning, facilitating efficient and accurate SED with significantly reduced model complexity. The studies on the DCASE 2023 Task 4A database showed that our proposed system having only one-third of the baseline model's parameters,  achieves superior performance in both PSDS1 and PSDS2 metrics. This validates the effectiveness of our approach, confirming its potential to develop lightweight yet robust SED systems suitable for resource-constrained edge devices.
\clearpage
\vfill\pagebreak
\footnotesize
\bibliographystyle{IEEEbib}
\bibliography{refs}
\end{document}